\documentclass{pasj00}
\draft

\begin{document}
\SetRunningHead{Meshalkina et al.}{Eruptions of Magnetic Ropes in
Two Homologous Solar Events}

\Received{???}
\Accepted{???}

\title{Eruptions of Magnetic Ropes in Two Homologous Solar Events of 2002 June 1 and 2:
a Key to Understanding an Enigmatic Flare}

 \author{Nataliya S. \textsc{Meshalkina}\altaffilmark{1,2},
 Arkadiy M. \textsc{Uralov}\altaffilmark{2},
 Victor V. \textsc{Grechnev}\altaffilmark{2},
 Alexander~T.~\textsc{Altyntsev}\altaffilmark{2},
 and Larisa~K.~\textsc{Kashapova}\altaffilmark{2}}
    \altaffiltext{1}{Nobeyama Radio Observatory, Minamimaki, Minamisaku, Nagano
       384-1305}\email{mtssnami@yandex.ru}
    \altaffiltext{2}{Institute of Solar-Terrestrial Physics, Lermontov St. 126A,
       Irkutsk 664033, Russia}

\KeyWords{Sun: filaments --- Sun: flares --- Sun: radio radiation
--- Sun: UV radiation}

\maketitle

\begin{abstract}

The goal of this paper is to understand the drivers, configurations,
and scenarios of two similar eruptive events, which occurred in
the same solar active region 9973 on 2002 June 1 and 2. The June 2
event was previously studied by Sui, Holman, and Dennis (2006,
2008), who concluded that it was challenging for popular flare
models. Using multi-spectral data, we analyze a combination of the
two events. Each of the events exhibited an evolving cusp-like
feature. We have revealed that these apparent ``cusps'' were most
likely mimicked by twisted magnetic flux ropes, but unlikely to
be related to the inverted Y-like magnetic configuration in the standard
flare model. The ropes originated inside a funnel-like magnetic
domain whose base was bounded by an EUV ring structure, and the
top was associated with a coronal null point. The ropes appear to
be the major drivers for the events, but their rise was not
triggered by reconnection in the coronal null point. We propose a
scenario and a three-dimensional scheme for these events in which
the filament eruptions and flares were caused by interaction of
the ropes.

Online material: mpeg animations.
\end{abstract}

\section{Introduction}
 \label{S-Introduction}

Close association between eruptive phenomena and flares is well
known, but causal relations between them have not yet been well
understood. Several models have been proposed to describe the vast
variety of phenomena observed in flares and eruptions. However,
observations sometimes offer challenges for the existing models.
In particular, coronal magnetic configurations in some events
remain unknown.

For example, many properties of long-decay eruptive flares appear
to be described by the ``standard'' flare model sometimes also
referred to as ``CSHKP'' for its main contributors
\citep{Car64,Sturrock66,Hirayama74,Kopp76}. This model was
advanced in many later studies, e.g.,
\authorcite{Somov1992} (\yearcite{Somov1992,Somov2006});
\authorcite{Shibata95b}
(\yearcite{Shibata95a,Shibata95b,Shibata96,Shibata99});
\authorcite{Yokoyama1998} (\yearcite{Yokoyama1998,Yokoyama2001}),
etc. In this scenario, the major energy release occurs in magnetic
reconnection in a vertical current sheet after an eruption of a
magnetic filament, rope, or plasmoid. A range of probable causes
of the appearance and acceleration of the ejection has been
considered, from the emergence of a twisted magnetic flux rope from
below the photosphere up to a gradual breakout reconnection high
in the corona. The ejected plasmoid, in turn, eventually evolves
into a coronal mass ejection or its part. One of the attributes of
this model is the cusp region at the bottom of the current sheet.
Shrinking flare loops are considered to be formed in the cusp
region. Attempts to relate the standard model to observations of
flares sometimes stumble upon problems, especially when impulsive
flares are considered. For example, cusps occasionally appear in
impulsive flares, but too late, when the major plasma heating and
particle acceleration processes are almost completed (see, e.g.,
\cite{Grechnev2002, Grechnev2006SolSys, Sui2008}).

An intriguing impulsive eruptive flare which occurred on 2002 June
2 was presented by \authorcite{Sui2006}
(\yearcite{Sui2006,Sui2008}). The authors discovered several
puzzling facts in the morphology and sequence of events observed
in this flare. Multiple-loop interactions appeared to be the cause
of the flare. The authors considered the emerging flux model
\citep{Heyvaerts77} and the magnetic breakout model
\citep{Antiochos99} and found both of them to explain some changes
of the loop morphology in the flare; however, none of these models
was found to explain all observational facts.
\authorcite{Sui2006} (\yearcite{Sui2006}) characterized the situation as an ``enigma
of a flare involving multiple-loop interactions''. The enigma
consisted of the following problems. Some erupting features
resembled magnetic flux ropes, but their nature and role in the
events were unclear. One more problem was related to an apparent
dynamic formation of a cusp-like feature observed in the event
after the termination of the impulsive phase of the flare. 
The nature of the ``cusp'' remained unclear. The standard model did
not seem to explain the phenomena, but the data that 
\authorcite{Sui2006} (\yearcite{Sui2006,Sui2008}) employed were 
not decisive in choosing an alternative model.

To address these unresolved issues, we note that another, very
similar eruptive event occurred in the same active region (AR) 32
hours earlier, on June 1. Both the June 1 and 2 events exhibited
much the same manifestations: very short durations of $\approx 10$
minutes in soft X-rays (SXR), comparable durations and total
fluxes of hard X-ray (HXR) and microwave bursts, analogous
eruptions and cusp-like features observed in extreme ultraviolet
(EUV), and similar flare configurations of a comparable size seen
in EUV, SXR, and HXR.

Various data are available for the June 1 event, i.e., 195~\AA\
images from the Transition Region and Coronal Explorer (TRACE,
\cite{Handy1999}), microwave images from the Nobeyama
Radioheliograph (NoRH, \cite{Nakajima1994}), magnetograms from the
Michelson Doppler Imager (MDI, \cite{Scherrer1995}) and EUV images
from the Extreme-Ultraviolet Imaging Telescope (EIT,
\cite{Delab1995}) on SOHO, as well as X-ray data from the Reuven
Ramaty High-Energy Solar Spectroscopic Imager (RHESSI,
\cite{Lin2002}). Utilizing this comprehensive data set for the
June 1 event and using the striking similarity between the June 1
and 2 events, we endeavor to understand their ``enigma''. Our
advantages as compared to the study of \authorcite{Sui2006}
(\yearcite{Sui2006,Sui2008}) are the possibility to combine
findings from observations of both these events and the
availability of NoRH microwave images for the June 1 event.

The goal of our study is to reveal the scenario of the two
homologous June 1 and 2 events, their magnetic configurations, and
major drivers. Section~\ref{S-Observations} addresses the
observations. Section~\ref{S-Interpretation} summarizes the
observational results and outlines a combined scenario of these
eruptive events. Section~\ref{S-Conclusion} summarizes our
findings and conclusions.

\section{Observations and Analysis}
 \label{S-Observations}

\subsection{Summary of both events}
 \label{S-Summary_both}

\begin{figure*}
  \begin{center}
    \FigureFile(170mm,200mm){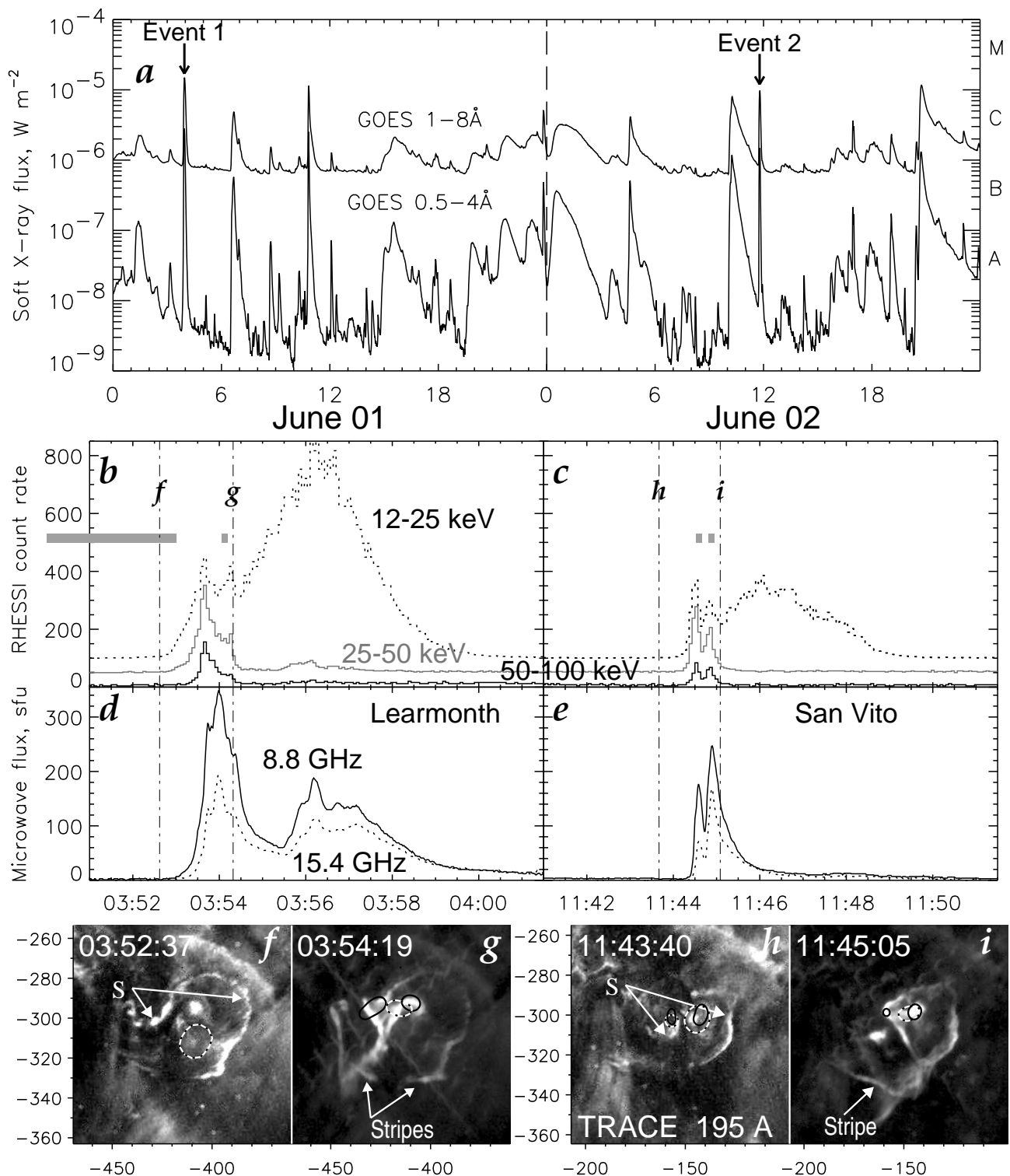}
  \end{center}
\caption{Summary of June 1 and 2 events. GOES SXR flux (a), RHESSI
count rates (b,c; vertically shifted for a better view), microwave
total flux (d,e), and TRACE 195~\AA\ images (hereafter, the axes
show the distance in arcsec from the solar disk center) just
before (f,h) and during the eruptions (g,i). Vertical broken lines
(b--e) mark the central times of the TRACE images. Contours (f--i)
are RHESSI images constructed with the Pixon method at 6--12~keV
(dotted; 3--12~keV in panel f) and 25--50~keV (black) over
intervals shown in panels b and c by gray bars. Contour levels are
40\% of the maximum brightness for event 1 and 30\% for event 2.
All images hereafter are shown nonlinearly to reveal the features
of interest.}
   \label{F-summary}
 \end{figure*}

Two similar impulsive eruptive events occurred in AR 9973
($\beta\gamma$-configuration) on 2002 June 1 and 2
(figure~\ref{F-summary}). The first event was associated with an
M1.5 flare (S19~E29) on June 1, which started at 03:50 (all times
hereafter are UT), peaked at 03:57, and ended at 04:01 according
to GOES reports. The second event was associated with a C9.4 flare
(S20~E09) on June 2 (11:41--11:47--11:50). Figure~\ref{F-summary}
shows some representative data related to both events (June 1
left, June 2 right). The scales are the same for both events.

Total duration of each event was $\approx\,$10 min. Both events
contained impulsive parts of $\sim\,$2 min followed by longer
thermal tails in low-energy RHESSI bands $< 25$~keV. The impulsive
parts had comparable fluxes in X-rays and microwaves, while HXR
and radio fluxes were somewhat weaker in the June 2 event. The
June 1 event contained the second enhancement from 03:55:30 to
03:59, which was rather weak in HXR at $>25$ keV, but strong in
microwaves. The corresponding second enhancement was weak in the
June 2 event.

We estimated the temperature and emission measure from two
SXR bands of GOES-10 following \citet{White2005}. In both events,
the average temperature reached about 15~MK at the main peak. For
the June 1 event, the total emission measure was $2 \times
10^{48}$~cm$^{-3}$ at the impulsive peak (03:54). With a size of
the SXR-emitting region of about 7 Mm (from a RHESSI 3--6~keV
image), the plasma density was $7 \times 10^{10}$~cm$^{-3}$ at
that time. The thermal microwave flux estimated for the June 1
event was $\leq 7$~sfu. Similar parameters were estimated for the
June 2 event. They suggest that the microwave bursts were almost
entirely non-thermal.

TRACE observed both events in the 195~\AA\ band with an interval
of 17~s. There were gaps in observations of the second event
during 11:45:44--11:55:06. The TRACE 195~\AA\ images in
figures~\ref{F-summary}f and \ref{F-summary}h show the pre-flare
configurations to be similar (S-like) in shape, and
figures~\ref{F-summary}g and \ref{F-summary}i show basically similar eruptions
of mutually wrapped structures. We produced TRACE 195~\AA\ movies
of either events TRACE\_195\_2002\_June\_01.mpg and
TRACE\_195\_2002\_June\_02.mpg, having nonlinearly processed the
images to reveal features of interest. The movies also show a
great deal of similarity between the two events.

Filaments existed in the active region before either event. In both
cases, S-like structures (marked ``S'' in
figures~\ref{F-summary}f and \ref{F-summary}h) resembling flare ribbons brightened up,
and then their eastern parts rose and erupted. Later on, their
remainder parts and the central regions became the footpoint areas
of the flare loops. During the events, stripes appeared adjoining
the S-like structures from the south and southeast as their
extensions (figures~\ref{F-summary}g and \ref{F-summary}i). All visible phenomena on
the solar surface were confined by the outermost boundary of the
S-like structure, the stripes, and footpoint areas of the flare
loops. We call this boundary a \textit{ring structure} (see double
contour in figure~\ref{F-TRACE_NoRH_June_1}e). This boundary
suggests that a nearly closed magnetic structure confined the
volume of the event. The cusp-like features as well as erupting
dark filaments and bright ropes were visible in both events. The
erupting filaments and ropes mutually wrapped, pushed their way
through the coronal configuration which confined them, and escaped
as a broad jet followed by outflowing dark material (surge). The
impulsive flares occurred at the same time. Using the similarity
between the two events, we combine findings from observations of
each events.

\subsection{The June 1 Event}
\label{S-June_1}

\subsubsection{The Configuration of the Flare Region}

\begin{figure}
  \begin{center}
    \FigureFile(67mm,132mm){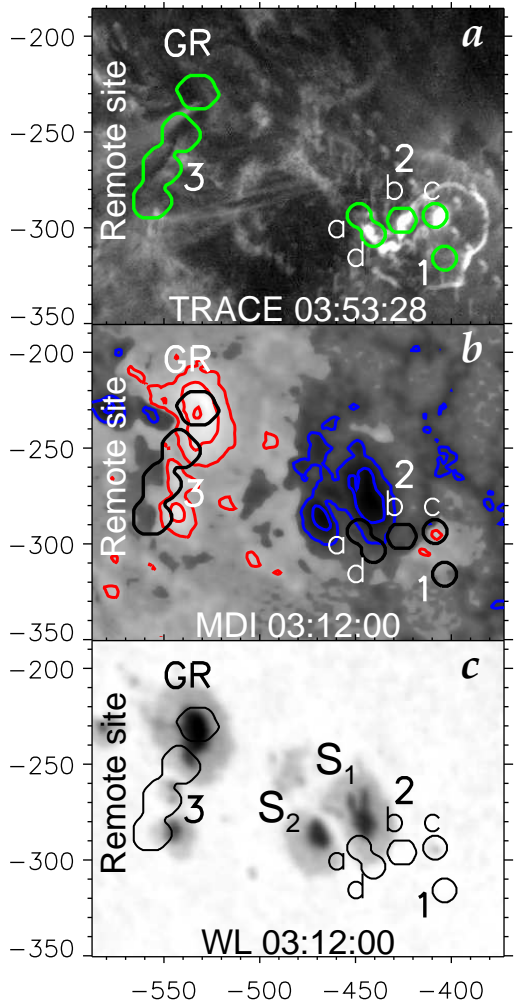}
  \end{center}
\caption{The June 1 event. Contours of 17~GHz flare sources 1--3
(green and black) superimposed on a TRACE 195~\AA\ image (a), an
MDI magnetogram (b), and an MDI white-light image (c). Light areas
and red contours represent N-polarity, dark areas and blue
contours show S-polarity (levels $\pm500\,\mathrm{G},
\pm1500\,\mathrm{G}$).}
   \label{F-TRACE_mag_radio}
 \end{figure}

The event of June 1 was imaged by the NoRH at 17~GHz (both Stokes
$I$ and $V$). During the flare, the NoRH beam size was
$13.3^{\prime \prime} \times 12.8^{\prime \prime}$. Microwave
images obtained during flares usually show a simple picture
consisting of a few major sources in which strong gyrosynchrotron
emission predominates. Nevertheless, microwave data carry
information about features which could be weaker, especially
prior to the main flaring, but important for understanding this
complex eruptive event. Figure~\ref{F-TRACE_mag_radio} shows a
TRACE 195~\AA\ image (a) and a SOHO/MDI magnetogram (b) as well as
white-light image (c) with superimposed contours of microwave
sources. The TRACE image in figure~\ref{F-TRACE_mag_radio}a shows
the ring structure consisting of the S-like configuration and two
southern stripes (indicated in figure~\ref{F-summary}g) before and
during the flare.

Two microwave-emitting regions labeled 1 and 2 were located within
the main flare site (see
figure~\ref{F-TRACE_NoRH_June_1-preflare}d), while it is possible
to discern four components a--d of region 2. Component 2a was in
between the S-polarity sunspots S$_1$ and S$_2$
(figure~\ref{F-TRACE_mag_radio}). Component 2c was located at a
small N-polarity sunspot. A 70\%-polarized steady gyroresonance
source labeled ``GR'' coincided with a large N-polarity sunspot. A
remote elongated region 3 was located south of it.

\subsubsection{TRACE 195~\AA\ and 17 GHz data}
 \label{TRACE_NoRH}

\begin{figure*}
   \begin{center}
     \FigureFile(170mm,46mm){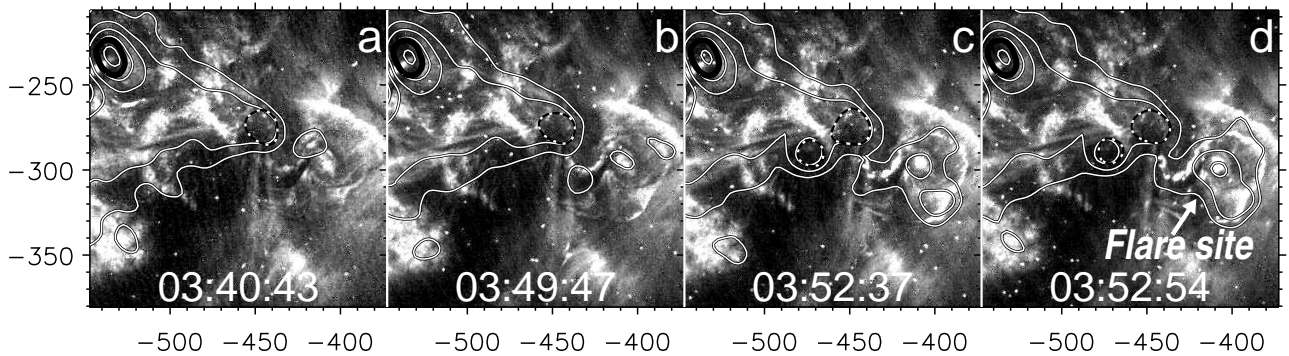}
   \end{center}
 \caption{TRACE 195~\AA\ and 17 GHz images of the June
 1 event over the pre-flare phase. The white solid contours
 (levels of $[12.5, 25, 50, 100, 200]\times 10^3$~K) show the total
 intensity. The black (in the upper left corner) and dotted
 contours show the positive and negative polarization respectively (50\%
 of the maximum of the absolute value). Imaging intervals for both
 TRACE and NoRH are 10~s around the specified times.}

    \label{F-TRACE_NoRH_June_1-preflare}
  \end{figure*}

\begin{figure*}
  \begin{center}
    \FigureFile(158mm,200mm){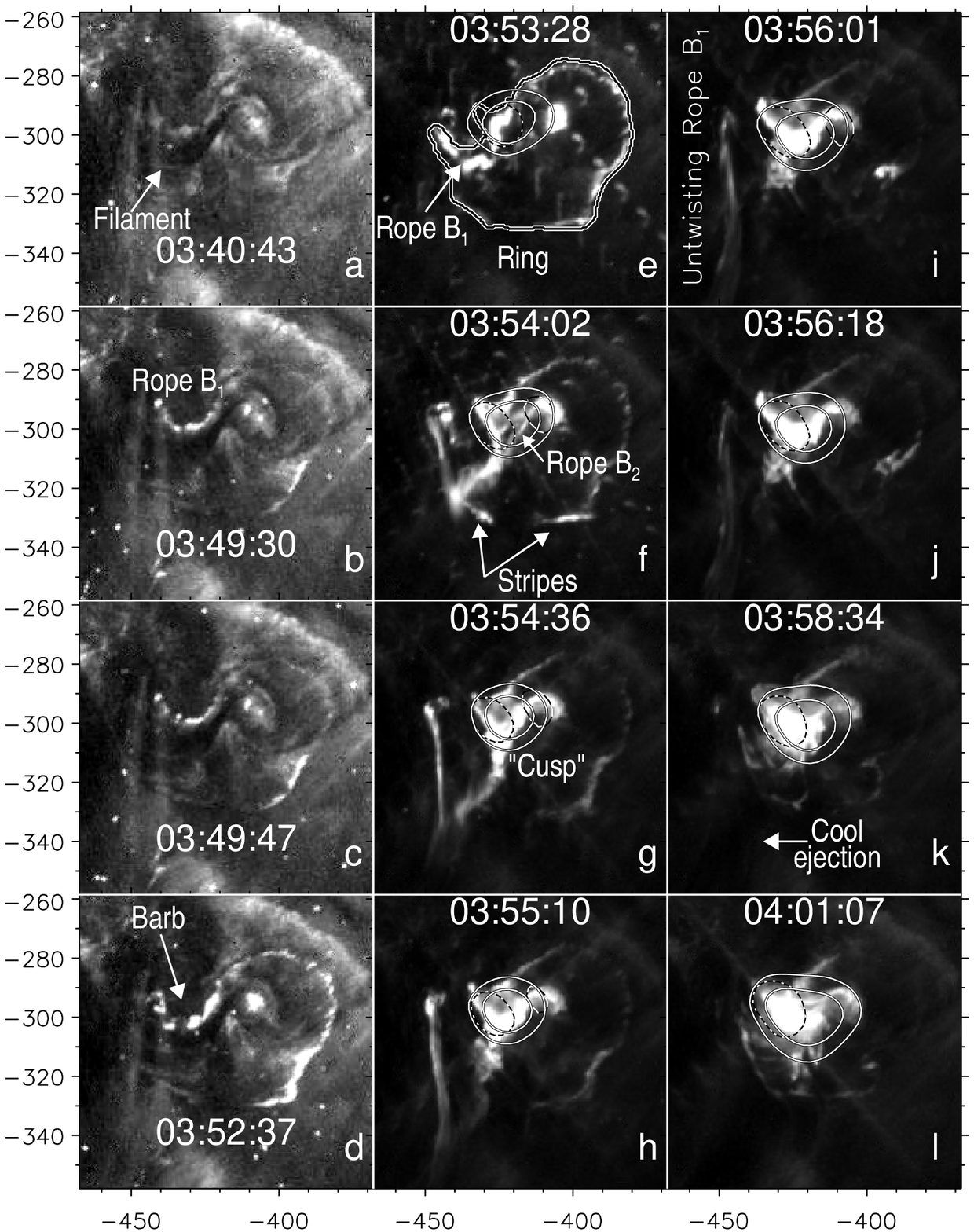}
  \end{center}
\caption{TRACE 195~\AA\ and simultaneous 17 GHz images of the June
1 event. The white solid contours show 20\% and 50\% of the
maximum intensity. The broken contours show the polarization
(Stokes $V$, dotted negative, dashed positive, both 50\%).
Contours of the 17 GHz images in the previous figure are
not shown in the left column. The double contour in panel (e)
shows the outer boundary of the ring structure. Slanted cross-like
dark/bright patterns in TRACE images are due to interference from
bright features on the CCD detector.}
   \label{F-TRACE_NoRH_June_1}
 \end{figure*}

Figure~\ref{F-TRACE_NoRH_June_1-preflare} shows the pre-flare
situation, and figure~\ref{F-TRACE_NoRH_June_1} presents some
milestones of the entire event. The S-like structure appeared in
195~\AA\ images at 03:20:27 (figures~\ref{F-TRACE_NoRH_June_1-preflare} and
\ref{F-TRACE_NoRH_June_1}; earlier images are not shown). The
eastern part of the S-like structure erupted during the event. The
behavior of this feature suggested that it was a magnetic flux
rope (denoted B$_1$ in figure~\ref{F-TRACE_NoRH_June_1}b). A dark
filament was visible south of rope B$_1$. Some of the long, dark
filament barbs occulted the bright rope B$_1$
(figures~\ref{F-TRACE_NoRH_June_1}c and \ref{F-TRACE_NoRH_June_1}d); hence, the rope must be
located under the filament.

\begin{figure*}
  \begin{center}
    \FigureFile(159mm,140mm){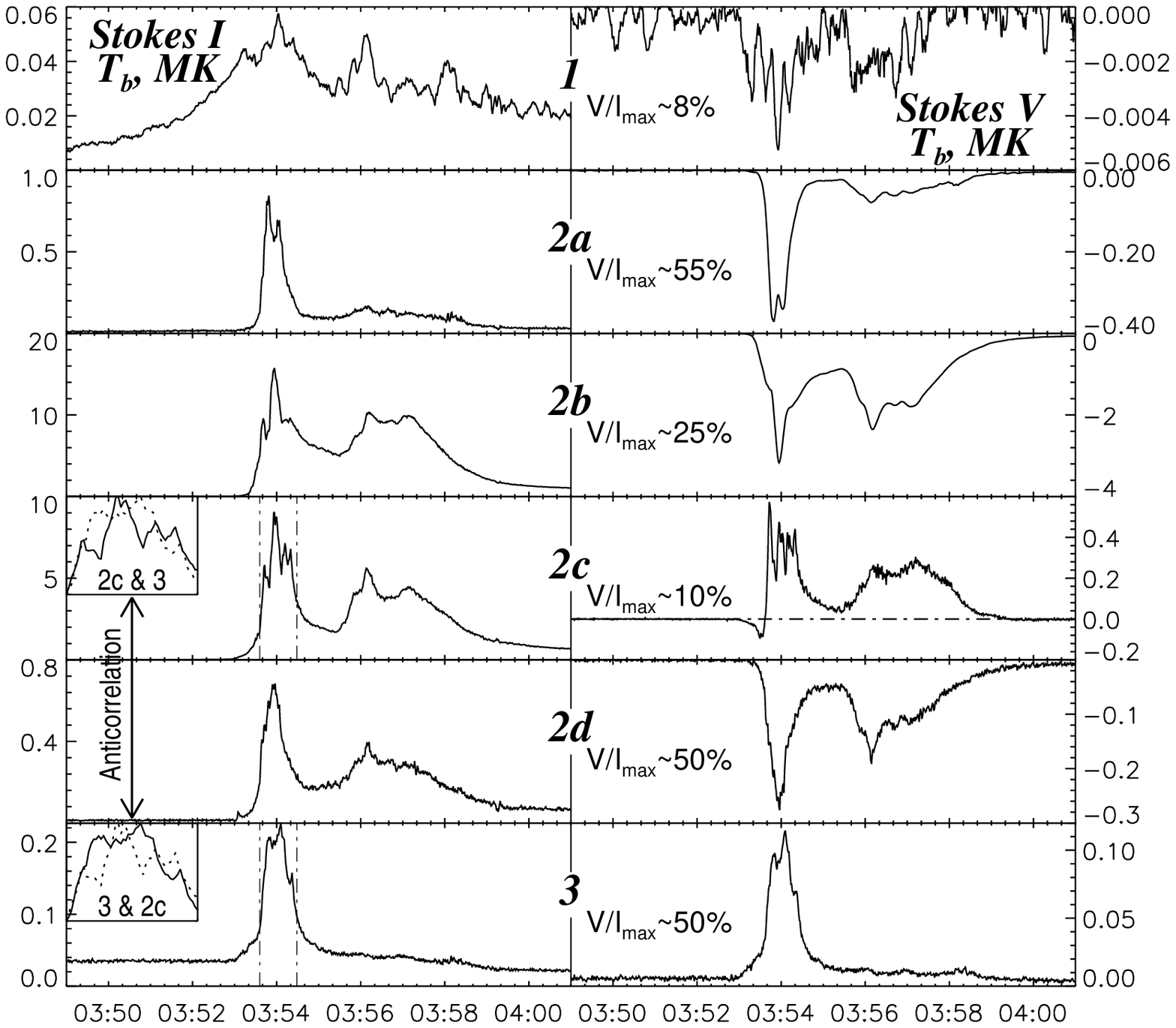}
  \end{center}
\caption{The June 1 event. Time profiles of the average brightness
temperature at 17 GHz in total intensity (left) and polarization
(right). "V/I$_{max}$" is the maximum degree of polarization in each
region.}
   \label{F-all_radio_timeprofs}
 \end{figure*}

At about 03:47, the western part of the S-structure brightened and
then started to expand outwards like a flare ribbon. Rope B$_1$
started to move southwards under the filament. The eruption
occurred at 03:53--03:55 (figures~\ref{F-TRACE_NoRH_June_1}e--h);
rope B$_1$ brightened, stretched, bent, carrying the dark filament
away, and finally disrupted. This interval corresponded to the
first microwave peak (see figure~\ref{F-all_radio_timeprofs}). Dark
outflow was afterwards seen in 195~\AA\ and H$\alpha$ images as a
surge.

We have revealed active flare sources at 17 GHz by analyzing the
variability of microwave images \citep{Grechnev2003,
Grechnev2006PASJ-acc}. The time profiles of the average brightness
temperature in total intensity and polarized emission for each
source are shown in figure~\ref{F-all_radio_timeprofs}.

In microwaves, region 1 brightened before the flare (figures
\ref{F-TRACE_mag_radio} and \ref{F-all_radio_timeprofs}). Its
gradual time profile suggests its mainly thermal nature.
Components a--d of region 2 were almost merged in total intensity,
but distinct in polarized emission. Region 2a made a smaller
contribution to the second peak than did regions 2b--d (the main
Stokes $I$ source slightly shifted southwards during the second peak).
The similarity of the time profiles of sources 2a--d in the main
flare site (figure~\ref{F-TRACE_NoRH_June_1-preflare}d) and the
remote source 3 indicates a connection between them during the
first peak. Their connection is additionally supported by the
anticorrelation of subsidiary peaks on top of the first main peak
in the time profiles of regions 2c and 3 (see insets in
figure~\ref{F-all_radio_timeprofs}). The absence of manifestations
of the second main peak in region 3 suggests disconnection of
regions 2 and 3 in the course of the event (cf.
\cite{Kundu2001,Grechnev_et_al2003}). A high degree of
polarization observed in these regions supports their non-thermal
nature.

\subsubsection{Large-Scale Coronal Disturbances}
 \label{S-Disturbances}

\begin{figure}
  \begin{center}
    \FigureFile(75mm,75mm){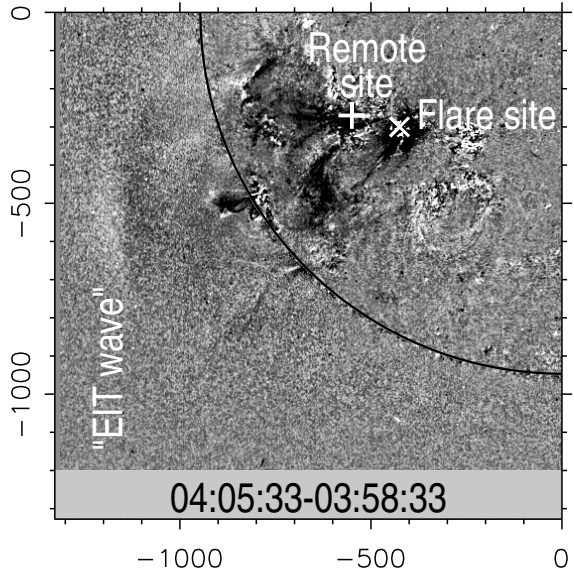}
  \end{center}
\caption{``EIT wave'' and dimmings in a SOHO/EIT 195~\AA\
difference image of the June 1 event.}
   \label{F-EIT}
 \end{figure}

Figure~\ref{F-EIT} shows coronal disturbances visible in a
difference image composed by subtracting SOHO/EIT 195~\AA\
images (03:58:33 and 04:05:33), both reprojected to 03:53:33
(close to the flare peak). The difference image shows an off-limb
``EIT wave'' and dimmings. The ``EIT wave'' was most likely due to
a coronal shock excited by the eruption (see \cite{Grechnev2008}).
Neither the ``EIT wave'' nor dimmings were detected in NoRH
images. The dimmings were located at the main flare site as well
as the remote site highlighted by the microwave source 3, and also
traced long loops between these sites. These loops were visible in
TRACE and EIT images before the eruption and disappeared
afterwards. 

This fact confirms that their eruption and, together with a 
probable disconnection between the radio source 2c and the former 
source 3, suggests a change in the magnetic connectivity between
the main and the remote sites.

Our TRACE\_195\_2002\_June\_01.mpg movie shows the
following phenomena during the microwave second main peak
(03:55:30--03:58:20): (i)~the mutually twisted erupting structures
transformed into a broad jet escaping from a confining coronal
magnetic configuration; (ii)~a wave-like disturbance propagated along the
ring structure counter-clockwise, and (iii)~weak threadlike
brightenings flickered near the remote site concurrently with
the wave-like disturbance as its counterpart. All these phenomena
suggest a possible reconnection between the escaping ejection and
the surrounding magnetic field (see Section~\ref{S-Flare_Cusp}).

\subsection{The June 2 Event}
 \label{S-June_2}

\begin{figure*}
  \begin{center}
    \FigureFile(158mm,200mm){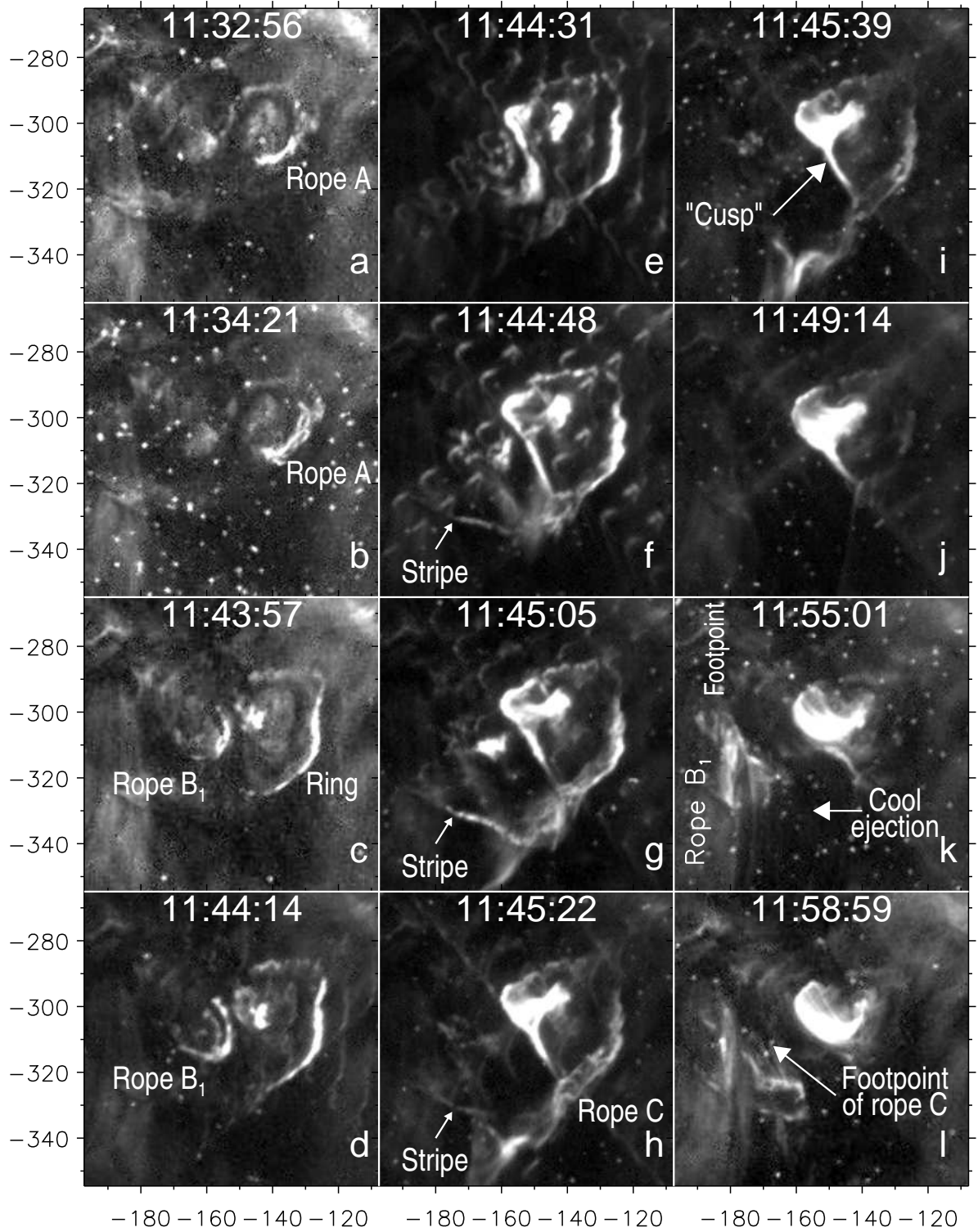}
  \end{center}
\caption{TRACE 195~\AA\ images of the June 2 event.}
   \label{F-TRACE_June_2}
 \end{figure*}

The overall story of the June 2 event previously addressed by
\authorcite{Sui2006} (\yearcite{Sui2006, Sui2008}) was basically
similar to the June 1 event. Some of its images observed with
TRACE at 195~\AA\ are shown in figure~\ref{F-TRACE_June_2}, and a
movie is available in the electronic version of this paper. We
processed these images heavier than did \authorcite{Sui2006} (\yearcite{Sui2006}),
which allowed us to reveal important features.

A bright twisted rising feature visible in
figures~\ref{F-TRACE_June_2}a and \ref{F-TRACE_June_2}b, which we
call rope A, was filled with material probably at a coronal
temperature. It appeared well before the event and disappeared
after 11:35. Most likely, the rope heated up (its temperature left
the sensitivity range of the 195~\AA\ channel) and erupted. A ring
structure came into view later on.

The event started at about 11:44 with a brightening of the S-like
structure (like the June 1 event). Its eastern part (rope B$_1$)
expanded southwest (figures~\ref{F-TRACE_June_2}c--f). The western
part moved west, and an erupting feature (presumably, yet another rope,
which we call rope C) showed up in its structure
(figures~\ref{F-TRACE_June_2}e--i). We did not see an analogous
feature in the previous event. Figures~\ref{F-TRACE_June_2}g--i and
a movie TRACE\_195\_2002\_June\_02.mpg display a collision of
ropes B$_1$ and C to produce a combined ejection transforming into
a broad outflowing jet. Rope B$_1$ turned southeast after the
collision, i.e., by about $90^{\circ}$ (figures
\ref{F-TRACE_June_2}f and \ref{F-TRACE_June_2}g), as if its
expansion were confined by a funnel.

The TRACE images were distorted due to interference on the CCD
detector, so that it is difficult to distinguish real features
from the interference fringe. Nevertheless, signatures of
untwisting ropes are detectable. Dark outflowing material (a
surge) is also visible in figures~\ref{F-TRACE_June_2}g--l.

\subsection{Evolution of Active Region 9973}

\begin{figure*}
  \begin{center}
    \FigureFile(170mm,113mm){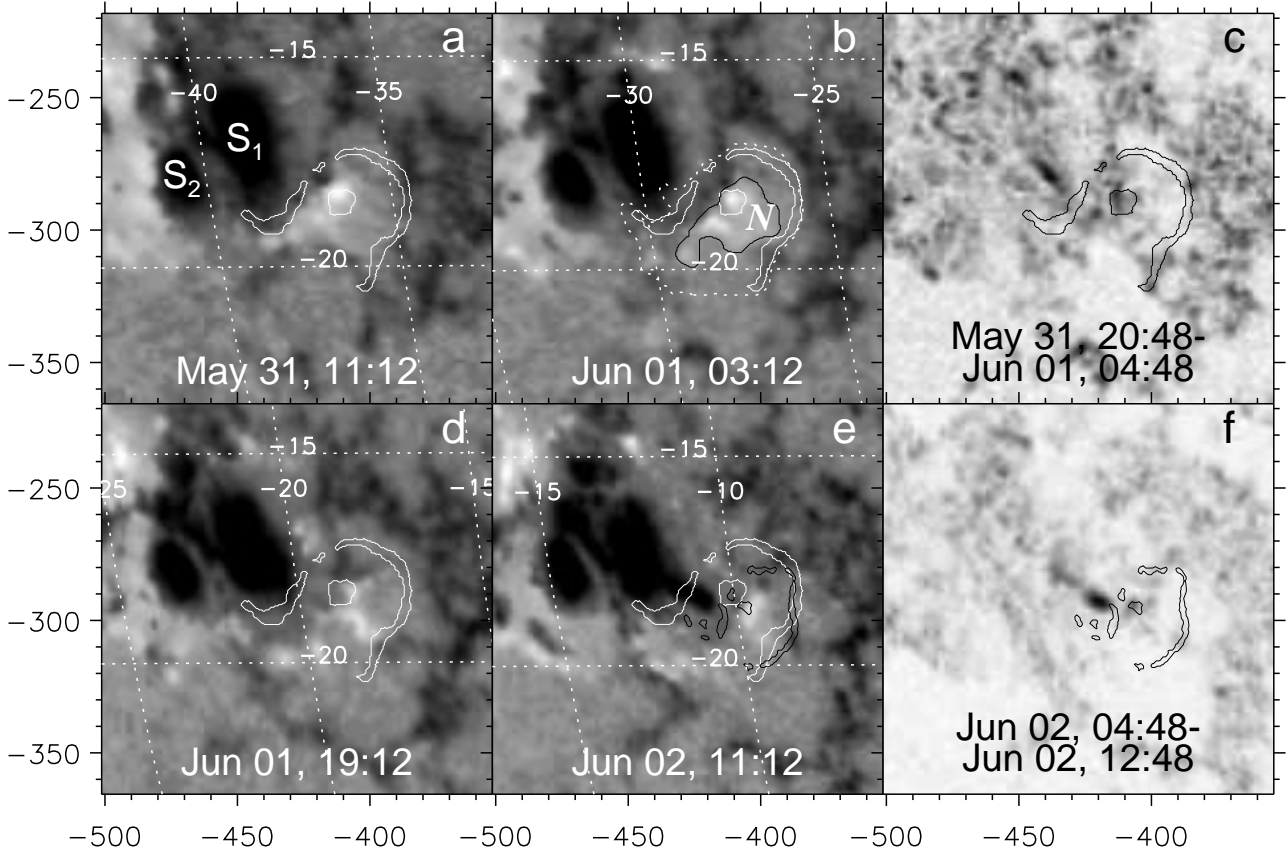}
  \end{center}
\caption{Evolution of AR~9973. Panels a,b, and d,e show SOHO/MDI
magnetograms; panels c,f show the negatives of variance maps
produced from the magnetograms over specified intervals. The
S-like structure visible in TRACE 195~\AA\ images is shown by the
white solid contours for June 1 (black in panel c) and black
contours for June 2 (in panels e, f). The white dotted contour in
panel b shows the ring structure, and the black contour in this
panel delineates the N-polarity island. All the images are
re-projected to the June 1 event. The white dotted grid (a--e)
shows the actual heliographic coordinates for each image.}
   \label{F-MDI}
 \end{figure*}

Similarity between the June 1 and 2 events implies the persistence
of the main photospheric configuration and reproduction of
pre-event conditions. Figure~\ref{F-MDI} presents the evolution of
AR~9973 shown by MDI magnetograms (see also the movie MDI.mpg).
Figures \ref{F-MDI}a and \ref{F-MDI}b, and \ref{F-MDI}d and \ref{F-MDI}e show the magnetograms.
Figures \ref{F-MDI}c and \ref{F-MDI}f show the negatives of
variance maps computed from the magnetograms over intervals
including the events (specified in the panels). Each variance map
(\cite{Grechnev2003}) was formed by calculating the variance
for each pixel in a set of magnetograms along their temporal
dimension. The most variable regions are seen in figures
\ref{F-MDI}c and \ref{F-MDI}f as the darkest patches. The major
changes were the development of sunspot S$_1$, the diminishing of the
N-polarity area (black contour in figure~\ref{F-MDI}b) inside the
ring structure (white dotted contour), and motions of magnetic
elements near the boundary of a supergranular cell.

The development of sunspot S$_1$ was manifest in increased field
strength in its southwestern part and a later growth of a narrow
extension which stressed the eastern part of the S-like structure
(figure~\ref{F-MDI}e). The small N-polarity area diminished and
eventually disappeared on June 3. The environment of the
S-structure (the boundary of the supergranular cell) exhibited
variations, while the western part of the S-structure was within a
relatively quiet area and remained almost unchanged. Thus, the
evolution of the active region explains the homology of the June 1
and 2 events and their differences.

\section{Summary and Interpretation}
 \label{S-Interpretation}

Using the resemblance of the June~1 and June~2 events, we combine
all observational findings in both these events, as if they were a
single event.

\subsection{Outline of the Coronal Magnetic Configuration}
 \label{S-Configuration}

\begin{figure*}
   \begin{center}
  \FigureFile(120mm,102mm){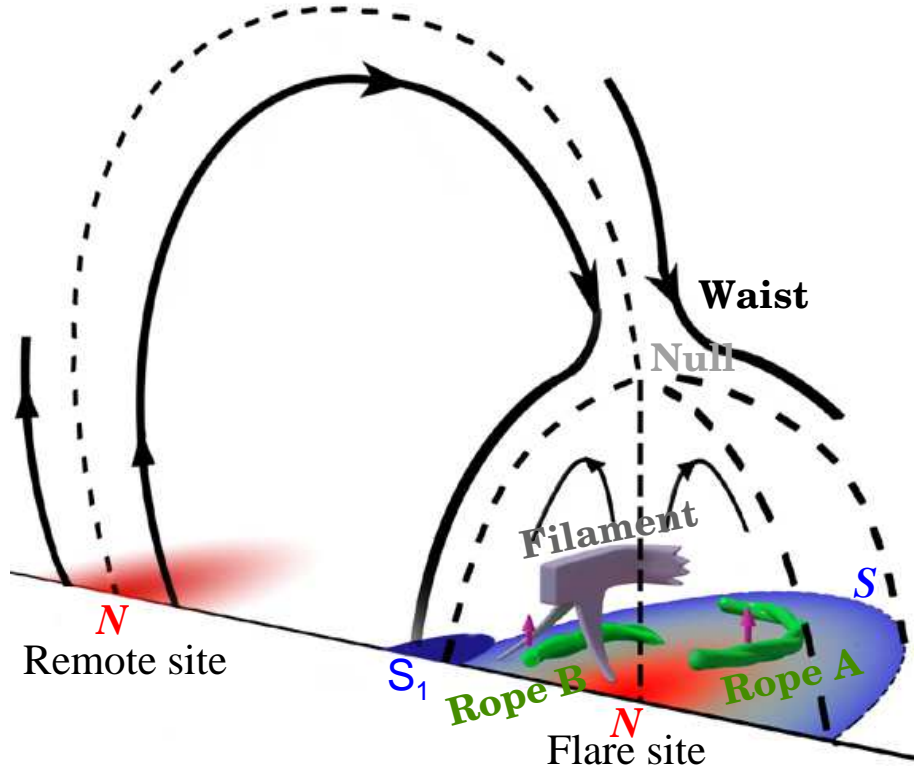}
   \end{center}
 \caption{The situation before the June 1 and
2 flares.  The N-polarity is red, the S-polarity is blue, the
neutral area is gray). Dashed lines denote separatrices, the
separatrix surface and its intersection with the photosphere. A
coronal null point is at the top of the volume in which the events
occurred. The remote site corresponds to the microwave source 3.
The frontal surface of the figure corresponds to a cross section
denoted by the arrows in figure~\ref{F-ropes_horizontal}. Rope A
rose and disappeared first. Next, rope B started to rise below the
filament (gray), which remained motionless.
        }
    \label{F-configuration}
  \end{figure*}

The main magnetic flux of AR~9973 was confined within a bipolar
structure consisting of an eastern N-polarity area and a western
S-polarity area (see figure~\ref{F-TRACE_mag_radio}b). The
photospheric region of the volume in which the eruptive events
occurred was encompassed by the ring structure (the white dotted
contour in figure~\ref{F-MDI}b), suggesting its correspondence to
the separatrix base (the dashed semi--oval in
figure~\ref{F-configuration} and the dashed circle in
figure~\ref{F-ropes_horizontal}). The ring structure was located
inside the extensive western S-polarity area and encompassed an
N-polarity island (black contour in figure~\ref{F-MDI}b). The
total fluxes of the N- and S-polarity magnetic fields calculated
inside the ring structure separately for the June 1 event were
approximately equal to each other (imbalance $<10\%$). This fact
favors an assumption that the magnetic flux of the N-polarity
island was mainly closed within the volume where the events
occurred. An additional support to this assumption is provided by
the fact that all observed motions of the erupting features in
either event were confined within a ``funnel''.

The ring structure looked like a flare ribbon occulted by a
multitude of dark absorbing features crossing it (see the western
part of the ring structure in figures~\ref{F-TRACE_NoRH_June_1}c--h
and its southern part in figure~\ref{F-TRACE_June_2}g labeled
``Stripe''). We are aware of a single paper describing a similar
phenomenon \citep{BorovikMyachin2002}. Some parts of the ring
structure resembled a quasi-stationary ``EIT wave'' stopping at
the separatrix surface. Such an ``EIT wave'' might be due to a
successive stretching of closed field lines (e.g.,
\cite{Chen2005}) driven by erupting flux ropes. The concept of a
separatrix surface is an idealization; instead, a quasi-separatrix
layer \citep{Demoulin1996} appears to match a realistic situation
better. The intersection of such a layer with the base of the
corona should be a set of stripes, which resembles the appearance
of the ring structure.

Our considerations lead to a configuration of the equilibrium
coronal magnetic field shown in figure~\ref{F-configuration} with
a coronal null point at its top (cf. \cite{Filippov1999};
\cite{GaryMoore2004}). The presence of a null point implies a
possibility of magnetic reconnection high in the corona. A
subsequent analysis, however, does {\it NOT} show the role of magnetic
reconnection in the coronal null point to be a crucial factor in
the initiation of the June 1 and 2 eruptive events.

\subsection{Magnetic Flux Ropes}
\label{S-Ropes}

\begin{figure}
   \begin{center}
  \FigureFile(75mm,75mm){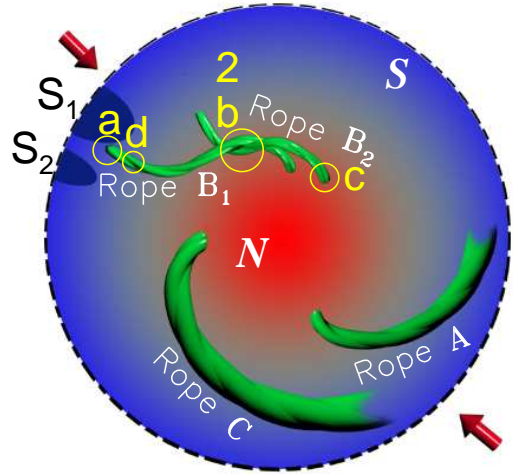}
   \end{center}
 \caption{Top view of figure~\ref{F-configuration}:
 probable positions of magnetic flux ropes (green). The area is bounded by
 the intersection of the separatrix surface with the photosphere
 (the outer dashed circle). The background shows the magnetic
 polarity at the photosphere. Ropes B$_1$ and B$_2$
 are components of rope B. The dark filament above is not shown.
 Open ends of ropes A and C
 indicate that the positions of their footpoints are not known.
 All the ropes are probably inter-related. Yellow circles denote
 microwave flare sources.
        }
    \label{F-ropes_horizontal}
  \end{figure}

According to \citet{Lites2005}, the presence of magnetic flux
ropes ``may be rather common in normal (i.e., non-$\delta$-type)
active regions'', and several flux ropes appeared to be involved
in the events. We call them ropes A, B, and C according to their
sequence in the ``combined event'' of June 1 and June 2.
Figure~\ref{F-configuration} outlines the pre-flare positions of
ropes A and B as well as the filament inside the volume enclosed
by the separatrix surface. The outer edge of the red/blue
semi-oval shows its intersection with the photosphere.
Figure~\ref{F-ropes_horizontal} shows the projection visible from
above (without the filament). The twist in the eastern footpoint
of rope B$_1$ visible in figure~\ref{F-TRACE_June_2}k means that
the electric current and the magnetic field are parallel. The same
direction of twist is assumed for ropes A and C.

Rope A appeared (see figures~\ref{F-TRACE_June_2}a and
\ref{F-TRACE_June_2}b) 8 min before the onset of the SXR flare
(GOES). The position of its visible southern footpoint
corresponded to the N-polarity. After its eruption, the ring
structure (figures~\ref{F-TRACE_June_2}c and later frames) appeared
simultaneously with the onset of the event in SXR. A probable
nature of the ring structure was discussed in
Section~\ref{S-Configuration}.

Rope B could appear (and start to move afterwards) in the course
of evolution of a low magnetic structure due to photospheric
motions and magnetic reconnection. The brightening of the rope
before its eruption could also be due to reconnection. To
co-ordinate our discussion with observations, we consider rope B
to consist of two interconnected ropes B$_1$ and B$_2$
(figures~\ref{F-configuration},\ref{F-ropes_horizontal} and
\ref{F-TRACE_NoRH_June_1},\ref{F-TRACE_June_2}). In the June 1
event, rope B$_1$ appeared under a dark filament like a bright
ribbon visible between barbs. This rope showed a large twist of
magnetic fields lines near its eastern footpoint F$_{B1S}$
(figures \ref{F-eruption}, \ref{F-TRACE_NoRH_June_1}i and \ref{F-TRACE_NoRH_June_1}j,
\ref{F-TRACE_June_2}k and \ref{F-TRACE_June_2}l, and movies). As the rope stretched, the
twist decreased. Rope B$_2$ manifested itself during the flare
(figures~\ref{F-TRACE_NoRH_June_1}f and
\ref{F-TRACE_NoRH_June_1}j).

One cannot rule out a possible emergence of ropes B$_1$ and B$_2$
from below the photosphere. Another possibility to propel rope
B$_1$ was an extra twisting of its eastern footpoint F$_{B1S}$.
Photospheric horizontal flows seen in the periphery of 
sunspots S1 and S2 appear to collide
near this footpoint (see the attached MDI movie). The history of rope B$_2$
is uncertain. Its presence was indicated by the cusp-like features
and suggested by the location of the main flare site in both
events being separated from a slow magnetic reconnection between ropes B$_1$ and B$_2$ probably started before the flare. Later, the ropes entered the fast flare reconnection stage and was combined into an interacting loop system which we call rope B.

Rope C (figure~\ref{F-TRACE_June_2}h) roughly corresponds to the
blue rope in figure~7 (lower middle panel) of \citet{Sui2006}. A
possible position of its eastern footpoint is shown in
figure~\ref{F-TRACE_June_2}l. Rotating motion of the rope was
visible, but its direction here remains unclear.

We assume the magnetic configuration consisting of flux ropes and
overlying magnetic field confining them to be metastable (see,
e.g., \cite{Sturrock2001}, \cite{Rachmeler2008}), and the toroidal
forces (\cite{Shafranov1966}, \cite{Chen1989},
\cite{Garren_Chen1994}) in rope B$_1$ to be the major driver of
the eruption. A system is metastable if it is stable (due to
confining fields) against small perturbations, but unstable to
sufficiently large perturbations (e.g., displacement, additional
twist, or plasma pressure increase). Therefore, an impulsive release
of magnetic energy is possible in such a situation. This can lead to
the ejection of the rope. If the amount of twist in it was 
sufficiently large, the rope was 
able to protrude like a hernia through the confining magnetic field
and escape, while the overlying field lines slid, without breaking, 
towards the anchored ends of the rope.
The assumption that the twist of rope B$_1$ was
really large is supported by the inclination angle of field lines
in its eastern leg (figure~\ref{F-TRACE_June_2}k) which shows that
the poloidal component of the magnetic field $B_\mathrm{p}$ was
nearly equal to the toroidal component $B_\mathrm{t}$ there. With
the observed ratio of the length to radius of rope B$_1$ being
$L/r \geq 5\pi$ (see figures~\ref{F-TRACE_NoRH_June_1}b and 
\ref{F-TRACE_NoRH_June_1}c and
\ref{F-TRACE_June_2}d), the equality of $B_\mathrm{t}$ and
$B_\mathrm{p}$ gives the amount of twist in rope B$_1$ to be $L/(2\pi r) B_\mathrm{p}/B_\mathrm{t} \geq 2.5$. Such a total
twist provides sufficient energy to drive an eruption (see, e.g.,
\cite{Sturrock2001}).

\subsection{Flare, Ejecta, and a Cusp-like Feature}
 \label{S-Flare_Cusp}

\begin{figure}
   \begin{center}
  \FigureFile(75mm,102mm){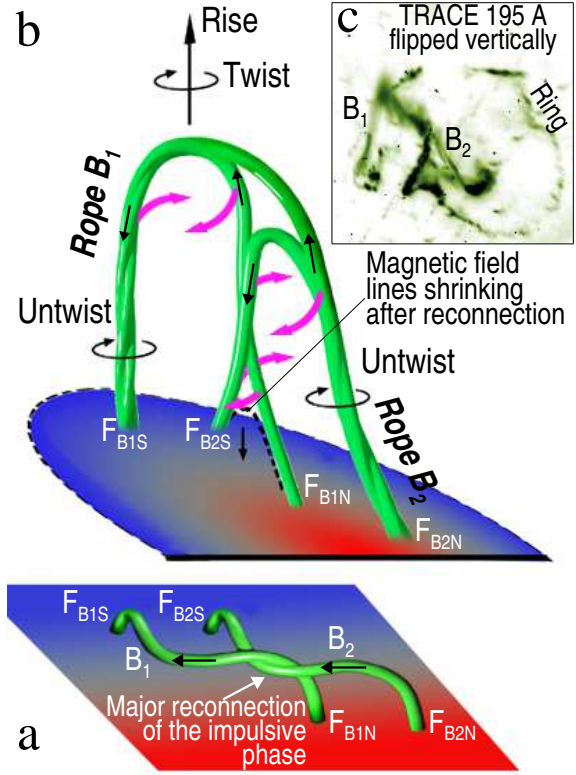}
   \end{center}
 \caption{Partial magnetic reconnection of ropes B$_1$ and B$_2$, their joint
rise, and rotation. The black arrows on the ropes indicate the
directions of electric currents. a)~The initial position of the
ropes. b)~The situation after the impulsive phase. c)~A negative
of the TRACE 195~\AA\ image at 03:54:02 flipped vertically to
match the scheme.
        }
    \label{F-eruption}
  \end{figure}

Figure~\ref{F-eruption} shows a scheme of the interaction between
ropes B$_1$ and B$_2$. The overlying field lines which prevent the
expansion of the configuration are not shown. The directions of
the electric currents in the ropes are indicated by black arrows.
Footpoints F$_{\mathrm{B1S}}$/F$_{\mathrm{B1N}}$ of rope B$_1$ and
F$_{\mathrm{B2S}}$/F$_{\mathrm{B2N}}$ of rope B$_2$ are assumed to
be fixed during the flare. The concept of mutual interaction of
magnetic flux ropes and the rule of a helical kink of a
current-carrying magnetic loop (e.g., \cite{Uralov1990}) were used
in drawing this scheme. Figure~\ref{F-eruption}a outlines the
initial locations of the ropes in static condition.

Rope B$_1$ rose for 3 min under the static filament, and the
observations did not show any motion of rope B$_2$ at that time.
Once the magnetic reconnection between ropes B$_1$ and B$_2$ has
started, their mutual configuration will determine the maximum
rate of reconnection, because the parallel components of currents in
the ropes attract each other. When rope B$_1$ moved upward on one
side of the filament, cutting its barbs and destabilizing it, the
first main peak of the June 1 flare occurred (between the
situations shown in figures~\ref{F-eruption}a and
\ref{F-eruption}b). Note that the measured acceleration of rope
B$_1$ was maximum at the rising stage of the impulsive phase.

Figure~\ref{F-eruption}b shows the situation after the impulsive
phase. The ``Untwist'' arrows show an untwisting of the rising
ropes just above their footpoints. The untwisting of the eastern
leg of rope B$_1$ is visible in the attached two movies. The reasons
of the untwisting can be illustrated by considering a freely
expanding rope with fixed footpoints. The first reason is the
expansion of the rope itself by conserving both its poloidal
and toroidal magnetic fluxes (i.e., a decrease of the longitudinal
current in the rope). The poloidal magnetic component decreases,
and the toroidal component remains nearly constant just above a
footpoint. The ratio of these components decreases at this place
which looks like untwisting. The second reason is a helical bending of
the whole rope that is accompanied by the decrease of the winding
of the field lines inside the rope (as occurs in a kink
instability). The pink arrows in figure~\ref{F-eruption}b
indicate the directions of motion for different parts of ropes
B$_1$ and B$_2$ in their helical bending. The pair of the rising
ropes rotates (the ``Twist'' arrow). The interacting parts of the
ropes above F$_{\mathrm{B1N}}$ and F$_{\mathrm{B2S}}$ tend to wrap
around each other. The crossing point of the ropes mimics a cusp
structure (figures \ref{F-eruption}b and \ref{F-eruption}c). If
the interacting parts of ropes B$_1$ and B$_2$ were infinitely
long, then the direction of their mutual twisting would correspond
to the ``Twist'' arrow. However, the ropes stayed side-by-side 
in the cusp region. The dashed line in
Figure~\ref{F-eruption}b shows a field line (flare loop) shrinking
down from the crossing point.

The cusp-like features were observed in the TRACE 195~\AA\ channel
(figures~\ref{F-TRACE_NoRH_June_1}f--h,~\ref{F-TRACE_June_2}g--j).
It is sensitive to the Fe{\sc xii} (0.5--2~MK) and Fe{\sc xxiv}
(11--26~MK) emissions (\cite{Handy1999}). Since hot features are
known to be fuzzy (e.g., \cite{Tousey1973}, \cite{Sheeley2004}),
the fine thready structures of the ``cusps'' imply that they were
at normal coronal temperatures of about 1.5~MK. Conversely, the
cusp related to the reconnection site in the standard model
should be hotter [$> 3$~MK; see, e.g.,
\authorcite{Yokoyama1998} (\yearcite{Yokoyama1998,Yokoyama2001})],
and therefore the appearance of the observed ``cusps'' does not
match the expectation. Also, the SXR 3--12 keV source in
figure~\ref{F-summary}f probably related to the coronal null
point was observed on June 1 as early as 03:50--03:53, but the
``cusp'', which must be associated with the SXR source, was formed
later (see figures~\ref{F-TRACE_NoRH_June_1}e--g), being not
co-spatial with it. Thus, the ``cusps'' most likely had no
relation to the inverted Y-type configuration in the standard model, but
were mimicked by cooler magnetic ropes.

Reconnection went on after the impulsive phase, because the
magnetic fields at the top of the cusp-like structure became
nearly antiparallel due to the stretch of the ropes. However, the
reconnection rate declined because the electric currents in the
ropes became also antiparallel and repelled each other. Both these
factors resulted in intertwisting of the ropes (see also
\cite{Hansen2004}). Note that the dual-rope initiation of a flare
(figure~\ref{F-eruption}) resembles the scheme proposed by
\citet{Uralov2002} and \citet{Grechnev2006PASJ-eru}.

The dual-rope configuration in figure~\ref{F-eruption}b was
distorted due to its instability to a helical twist. Moreover, the
distortion of ropes B and C was also influenced by the outer
magnetic structure encompassing the volume in which the events
occurred. This volume resembles a ``funnel'' with a narrow waist
(see figure~\ref{F-configuration}). The lower part of this
``funnel'' is bounded by a hemispheric separatrix surface. The
optimal escape path of ejections runs through the waist. The
``magnetic funnel'' confines the ejections as suggested by the
movies. The interaction of the ejections with this ``funnel'' and
their pass through the waist appear to have produced a wave
disturbance in the corona observed as an ``EIT wave'' (see
Section~\ref{S-Disturbances}) rapidly  expanding above the limb.

The motion of the mutually wrapped dual-rope ejection through the
waist must be accompanied by reconnection of the magnetic field
lines of the ejection with those of, or near to, the separatrix
surface. A response to this process could be concurrent EUV
disturbances that propagate along the ring structure and near the
remote site (see Section~\ref{S-Disturbances} and
figure~\ref{F-configuration}).

\subsection{Microwave Emission in the June 1 Event}
 \label{S-Microwave_Data}

The microwave pre-flare emission was mainly thermal, as shown by
contours in figure~\ref{F-TRACE_NoRH_June_1-preflare}, and the
only compact microwave sources were related to the major sunspots.
The ring structure appeared in microwaves, too, similarly to EUV (see
figures \ref{F-TRACE_NoRH_June_1-preflare}c and
\ref{F-TRACE_NoRH_June_1-preflare}d), indicating heating. All
flaring microwave sources (except for source 1) in
figure~\ref{F-TRACE_mag_radio} were non-thermal (see
figure~\ref{F-all_radio_timeprofs}).

The flare sources 2a--d were located along rope B consisting of
ropes B$_1$ and B$_2$ (figure~\ref{F-ropes_horizontal}). Source 2a
was located in the eastern footpoint of rope B$_1$, where its
untwisting was observed. Source 2b presumably corresponded to the
intersection region of ropes B$_1$ and B$_2$. Source 2c was
associated with the western footpoint of rope B$_2$. The time
profiles in figure~\ref{F-all_radio_timeprofs} suggest the connections
between all of these flare sources. Their non-thermal nature and
localization in ropes B$_1$ and B$_2$ indicate reconnection of the
ropes, confirming the scheme in figure~\ref{F-eruption}.

The detailed correspondence of the time profiles of the non-thermal
sources 2 and 3 during the first main peak 
(see figure~\ref{F-TRACE_mag_radio}) suggests their connection. 
The long loops detected in the TRACE 195~\AA\ images
between sources 2 and 3 traced a magnetic channel connecting the
flare and the remote sites. Particles responsible for the remote
microwave source 3 could arrive from the main flare site through
a change in magnetic connectivity while the
ejection was passing through the waist (see Sections~\ref{S-Disturbances} and
\ref{S-Flare_Cusp}).

\subsection{About Soft X-ray Sources}

\authorcite{Sui2006} (\yearcite{Sui2006}, Section 2.3) have not
revealed in the June 2 event any X-ray emission along the western
part of the ring structure (which they called ribbon A).
Similarly, no X-ray emission from this region was detected in the
June 1 event. These facts confirm our conclusion that this feature
was not a flare ribbon. \citet{Sui2006} also noticed in the same
Section that the main SXR sources, as shown by RHESSI at
6--12~keV, corresponded to the footpoints in the June 2 event,
which was unusual (see also our figure~\ref{F-summary}h). To explain this
fact, \citet{Sui2006} suggested a low density and temperature in
SXR-emitting loops, but our estimates
(Section~\ref{S-Summary_both}) do not confirm these assumptions.
Another possibility could be an enhanced brightness of overlapping
legs of the loops visible at small angles to the line of sight.
The RHESSI SXR images in question temporally correspond to the
TRACE 195~\AA\ images in figures \ref{F-TRACE_June_2}e and
\ref{F-TRACE_June_2}f, which clearly show structures with
temperatures of 1--1.5~MK. Images of hotter ($>11$~MK) 
structures coming at the same time in the Fe{\sc xxiv} 
passband of the TRACE 195~\AA\ channel are expected to be 
darker and fuzzier than the 1--1.5~MK structures registered 
in its major Fe{\sc xii} passband (see, e.g., \cite{Tousey1973}, 
\cite{Sheeley2004}, \cite{Grechnev2006PASJ-plasma}). Hence, we 
would not be able to separate the hot and the cooler structures 
in the TRACE 195~\AA\ images. However, the loops emitting at 
6--12~keV are expected to appear at
195~\AA\ after their cooling, and we simulated the SXR image from
later TRACE 195~\AA\ frames by blurring them to imitate their
previous fuzziness. The result roughly resembles the RHESSI image
in question and demonstrates that one might deal here with an
effect of integration of the column emission measure explaining
the enhanced SXR brightness of the footpoint regions.

\subsection{Initiation of the Events and Flare Models}
 \label{S-Initiation}

The geometry of the phenomenon inspires one to assume that the
appearance of the ring structure and the eruptions were caused by
magnetic reconnection in the coronal null point (see
figure~\ref{F-configuration}). In the breakout model, it is the
first step initiating the eruption \citep{Antiochos99}. However,
observations in the present events did not show the role of the 
``breakout'' reconnection high in the corona to be crucial in 
the triggering and course of the events for the following reasons.

\begin{enumerate}

\item
 The ring structure appeared after, rather than before, the eruption of
rope A.

\item
 The outermost boundary of the ring structure was well-defined. Reconnection
in a null point would not imply such a boundary.

\item
 The rise of rope B started under the almost static filament, which
protected it from the influence of the ``breakout'' reconnection.

\end{enumerate}

On the other hand, the rise and eruption of rope A preceded the
onset of the event in SXR and EUV, and the eruption onset of rope
B preceded the flare in microwaves and HXR. These facts indicate
that reconnection in the lower corona (e.g., ``tether cutting'')
was not a major launching factor for the events. This outcome is
consistent with the conclusion of \citet{Sterling2001}, ``the
tether cutting reconnection may still be occurring, but only after
the eruption onset was triggered by some other process''.

Unlike the standard model, in which flare reconnection is
initiated by a rise of a single magnetic rope, we have discussed
the initiation of the eruptive flares by reconnection of two
rising ropes. This explains the location of the main flare site
separated from the central, fastest-moving part of rope B$_1$. The dual-rope
configuration mimicked the cusp, which had no relation to the hot cusp
in the standard model. Reconnection between the ropes explains
non-thermal radio emission from the eastern footpoint of rope B,
the driver of the flare.

\section{Conclusion}
 \label{S-Conclusion}

The combined multi-spectral analysis of two homologous eruptive
events of 2002 June 1 and 2 allowed us to outline their common
picture. Similarity of the photospheric conditions and persistent
photospheric flows determined the resemblance between the events.
The boundary of the sites where the events occurred was a
time-evolving structure visible in EUV as a ring.

The observations left the impression that it was not possible to
perceive the scenarios of the events by considering each of them
separately. We have combined all the observed facts from both these events,
as if they were a single event, and proposed their common
scenario. We have concluded that their major drivers were
eruptions of magnetic flux ropes. Neither the triggering of the
events nor their evolution appear to be controlled by processes at
the coronal null point. Such processes were probably present, but
they did not determine the initiation of the events. The standard
model, the breakout model, as well as other models employing
magnetic reconnection high in the corona do not help to
understand these impulsive events.

\authorcite{Sui2006} (\yearcite{Sui2006,Sui2008})
highlighted problems in the interpretation of the 2002 June 2 event in
terms of widely accepted theoretical schemes and emphasized their
importance. Interpretation of the two events made in our paper is
based on a scheme involving the eruption of interacting flux
ropes. The scheme is inherently three-dimensional and hardly
reducible to the 2- or 2.5-dimensional geometry while maintaining
its essence.

\medskip

{\em Acknowledgements.} We thank K. Shibasaki, V.F. Melnikov, S.M.
White, D.M. Prokof`ev and D.S. Meshalkin for fruitful discussions and assistance.
We thank the referees for useful recommendations. We are
especially indebted to the second referee for his attention and great
efforts to improve the paper. The results presented here have 
become possible due to the usage of microwave
imaging data obtained with the solar dedicated Nobeyama
Radioheliograph operating continuously for over a decade.
N.S.M., A.M.U., and V.V.G. wish to thank the staff of NRO/NAOJ
(Japan) for their help and hospitality. This study was supported
by the Russian project of RFBR No. 07-02-00101, 08-02-92204, 09-02-00226.

We are grateful to the instrumental teams of the NoRH,
TRACE, RHESSI, and SOHO mission for their
open-data policies. SOHO is a project of international cooperation
between ESA and NASA.

\end{document}